\documentclass[useAMS,usenatbib]{mn2e}
\usepackage{graphicx,rotating,subfigure,floatflt,amsmath,wasysym}
\usepackage{graphicx,subfigure,floatflt,amsmath,wasysym, amssymb}
\usepackage{here}
\usepackage{setspace}
\usepackage{lscape}
\title[Extragalactic extinction law]{Determining the extragalactic extinction law with SALT. II. Additional sample {\thanks{based on observations made with the Southern African Large Telescope (SALT) }}}

\author[Ido Finkelman et al.]{Ido Finkelman$^{1}$\thanks{E-mail:
ido@wise.tau.ac.il (IF); noah@wise.tau.ac.il (NB); akniazev@saao.ac.za (AYK);  petri@saao.ac.za (PV); dibnob@saao.ac.za (DAHB); dod@saao.ac.za (DOD); amanda@saao.ac.za (AG); hashimot@ntnu.edu.tw (YH); nsl@saao.ac.za (NL); erc@saao.ac.za (ERC); rrs@saao.ac.za (RS)}, Noah Brosch$^{1}$, Alexei Y. Kniazev$^{2,3}$, Petri V\"{a}is\"{a}nen$^{2,3}$, 
\newauthor David A.H. Buckley$^{2,3}$, Darragh O'Donoghue$^{2,3}$, Amanda Gulbis$^{2,3}$, Yas Hashimoto$^{4}$, 
\newauthor Nicola Loaring$^{2,3}$, Encarni Romero-Colmenero$^{2,3}$ and Ramotholo Sefako$^{2,3}$\\
$^{1}$The Wise Observatory and the Raymond and  Beverly Sackler School of Physics and
Astronomy, the Faculty of Exact
Sciences, \\ Tel Aviv University, Tel Aviv 69978, Israel\\
$^{2}$South African Astronomical Observatory, PO Box 9, 7935 Observatory, Cape Town, South Africa \\
$^{3}$Southern African Large Telescope Foundation, PO Box 9, 7935 Observatory, Cape Town, South Africa\\
$^{4}$Department of Earth Sciences, National Taiwan Normal University, No.88, Sec. 4, Tingzhou Rd., Wenshan District,\\ 
Taipei 11677, Taiwan R.O.C}
\begin{document}

\date{Accepted 2010 July 08. Received 2010 July 06; in original form 2010 June 14}

\pagerange{\pageref{firstpage}--\pageref{lastpage}} \pubyear{2002}

\maketitle

\label{firstpage}

\begin{abstract}
We present new results from an on-going programme to study the dust extragalactic extinction law in E/S0 galaxies with dust lanes with the Southern African Large Telescope (SALT) during its performance-verification phase. 
The wavelength dependence of the dust extinction for seven galaxies is derived in six spectral bands ranging from the near-ultraviolet atmospheric cutoff to the near-infrared.
The derivation of an extinction law is performed by fitting model galaxies to the unextinguished parts of the image in each spectral band, and subtracting from these the actual images.
We compare our results with the derived extinction law in the Galaxy and find them to run parallel to the Galactic extinction curve with a mean total-to-selective extinction value of $R_V=2.71 \pm 0.43$.
We use total optical extinction values to estimate the dust mass for each galaxy, compare these with dust masses derived from IRAS measurements, and find them to range from $10^4$ to $10^7M_{\odot}$. 
We study the case of the well-known dust-lane galaxy NGC2685 for which HST/WFPC2 data is available to test the dust distribution on different scales. Our results imply a scale-free dust distribution across the dust lanes, at least within $\sim 1\arcsec$ ($\sim60$ pc) regions.
\end{abstract}


\section*{Introduction}
Elliptical galaxies were considered for many years to be inert systems, devoid of interstellar matter. 
The study of absorbing material in early-type galaxies gained interest with the identification of a class of galaxies characterized by an elliptical-like stellar body crossed by dark lanes or showing dark patches (Bertola \& Galletta 1978; Bertola 1987).
These are believed to be light extinguishing dust clouds.
Much of our knowledge of the dust grain properties is derived from the analysis of the attenuation of starlight by the absorption and scattering of light by grains in dusty structures. In particular, deriving the normalized amount of extinction as a function of wavelength, also known as the extinction law, is a useful tool to study the dust grain size and chemical composition.

The extinction law in the Milky Way (MW) is usually studied by observing stars with similar spectral type and luminosity class but with different foreground extinction (Savage \& Mathis 1979; Cardelli, Clayton \& Mathis 1989). Since this method requires individual stars to be resolved, different methods must be invoked to study the extinction law even in nearby galaxies. In galaxies where the underlying light distribution across the stellar body varies smoothly, namely ellipticals and lenticulars, dust lanes and patches appear as local disturbances of the global brightness pattern. Masking the dusty regions allows the extraction of a dust-free model of the original underlying galaxy and the derivation of extinction at these regions. 

A significant number of E/S0 galaxies have been investigated with this method (see Goudfrooij et al.\ 1994; Patil et al.\ 2007; Finkelman et al.\ 2008; Finkelman et al.\ 2010). The overall result from these studies is that the extragalactic dust is amazingly similar to the dust in the MW, at least in the optical part of the spectrum. One possible interpretation of this result is that dust grains in different objects seem to have similar sizes and chemical composition. Such a similarity may have implications on the physical conditions and/or the origin of the dust at least for this class of galaxies. 

This paper is the second in our on-going program to observe dark-lane E/S0 galaxies with the Southern African Large Telescope (SALT; Buckley, Swart \& Meiring 2006) and its CCD camera (SALTICAM, O'Donoghue et al.\ 2006). 
We present our analysis of seven dust-lane E/S0 galaxies, adding to the nine objects recently studied by Finkelman et al.\ (2008).
Observing nearby dust-lane E/S0 with an (effectively) nine-meter telescope has the specific advantage of achieving reasonable sensitivity even at the shortest spectral band used, despite the atmospheric extinction. The extention of the optical study of the extinction law towards shorter wavelength allows the extraction of more information about the extragalactic extinction law close to the blue atmospheric cutoff where differences among extinction curves are better detected.

This paper is organized as follows: \S~\ref{S:Obs_and_Red}
gives a description of all the observations and data reduction; we present and analyze our results in \S~\ref{S:Analysis}, discuss them in \S~\ref{S:Discussion}, and summarize our conclusions in \S~\ref{S:summ}.

\section{Observations and data reduction}
\label{S:Obs_and_Red}
CCD imaging observations of 12 dust-lane E/S0 galaxies were performed in service mode from February to May 2009 with SALT during its Performance Verification (PV) phase.
The observations were performed using the Johnson-Cousins B, V, R \& I filters, as well as two short-wavelength interference filters, U1 and U2, with transmission bands peaking at 340 nm (FWHM 35 nm) and 380 nm (FWHM 40 nm) respectively.
The SALTICAM is a CCD mosaic of two 2048x4102 pixel CCDs. The full SALTICAM image is about 9'.6 by 9'.6 with pixels of $\sim$0".28 (after  on-chip binning by a factor of two), but the nominal science field is eight arcmin.

The sample galaxies are listed in Table \ref{t:Obs} with their coordinates, morphological classification and optical size  taken from the RC3 catalog (de Vaucouleurs et al.\ 1992) or from HyperLEDA (Paturel et al.\ 2003).
The objects were selected from catalogs of objects with dust lanes (Hawarden et al.\ 1981; Ebneter \& Balick 1985; Bertola 1987) and were chosen according to visibility conditions and availability of observing nights, and not as part of a complete sample.

The observations consisted of short-duration integrations while the filter wheel cycled between the six filters, resulting in six images per cycle. 
Typical exposure times were 200 sec for the U1 \& U2 bands, 30 sec for the B \& V bands, and 15 sec for the R \& I bands.
They were determined to be sufficiently short to prevent image trailing when operating in the SALTICAM unguided mode (but with SALT tracking) or overexposure of the brightest part of each galaxy.
Due to the limited image quality of the telescope during the PV phase the workable field was limited to about four arcmin. The area nevertheless was sufficient to obtain minimally overlapping multiple dithered images of the galaxies in the central regions of both CCDs. 

We created a sky flat-field from the set of exposures in each field for each filter by median-combining the scaled images without correcting for dithering. We then corrected the images for flat-field and created a count-frequency histogram for each galaxy surroundings. We fitted a Gaussian distribution curve to the peak of the histogram, where the most frequent count value was assumed to represent the sky background. The images where the histogram significantly deviated from a Gaussian shape were rejected from the analysis.
The reduced images were geometrically aligned by measuring centroids of several common stars in the galaxy frames and were then background-subtracted and combined to improve the S/N ratio. This alignment procedure involved {\small IRAF}\footnote{{\small IRAF} is distributed by the National Optical Astronomy Observatories (NOAO), which is operated by the Association of Universities, Inc. (AURA) under co-operative agreement with the National Science Foundation} tasks for translation and rotation of the images, so that a small amount of blurring was introduced, affecting the positional accuracy to a few hundredths of a pixel. 

\begin{figure*}
\begin{center}
\begin{tabular}{ccc}
 \includegraphics[width=6cm]{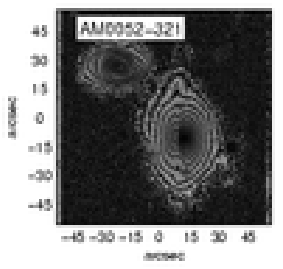} &  \includegraphics[width=6cm]{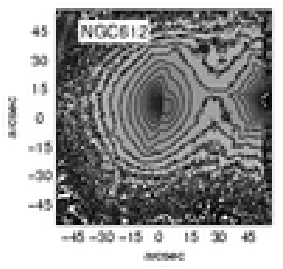} &\includegraphics[width=6cm]{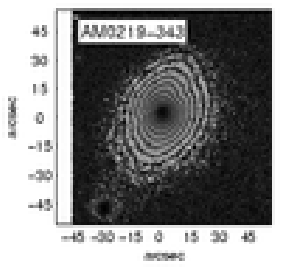}\\
  \vspace{-3mm}
 \includegraphics[width=6cm]{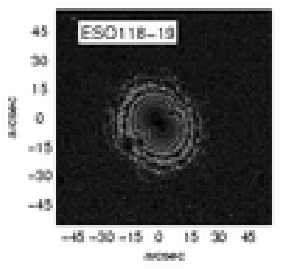} & \includegraphics[width=6cm]{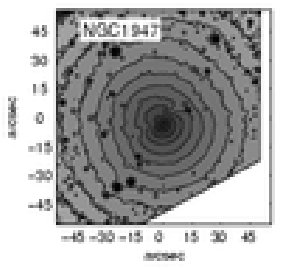} & \includegraphics[width=6cm]{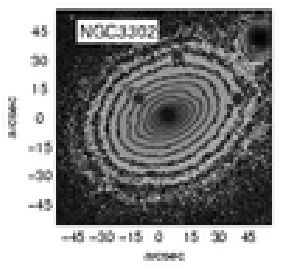}  \\
  \vspace{-3mm}
  \includegraphics[width=6cm]{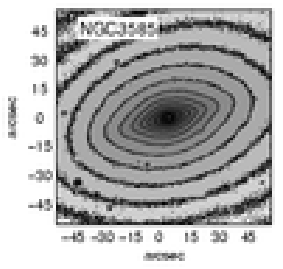} & \includegraphics[width=6cm]{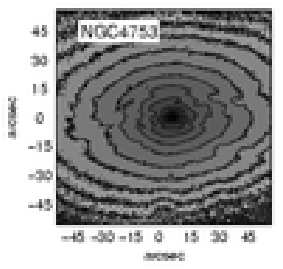} & \includegraphics[width=6cm]{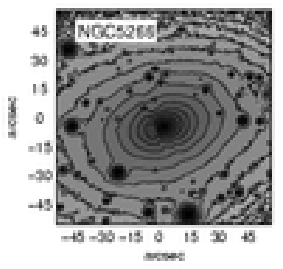}\\
 \vspace{-3mm}
\includegraphics[width=6cm]{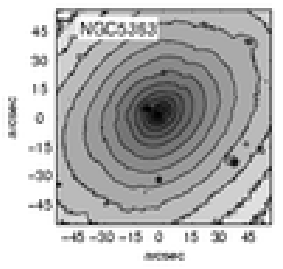} & \includegraphics[width=6cm]{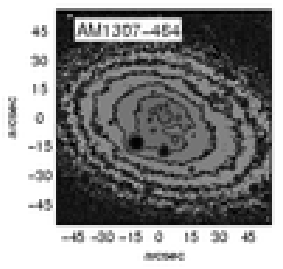} & \includegraphics[width=6cm]{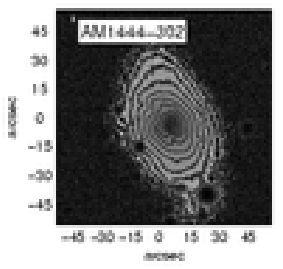} 
\end{tabular}
\end{center}
\caption{R-band contour maps.}
 \label{f:Rmaps}
\end{figure*}

\begin{figure*}
\begin{center}
\begin{tabular}{ccc}
 \includegraphics[width=6cm]{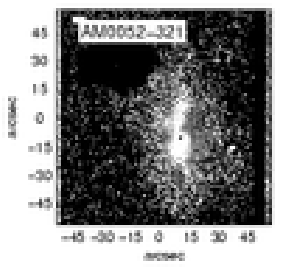} &  \includegraphics[width=6cm]{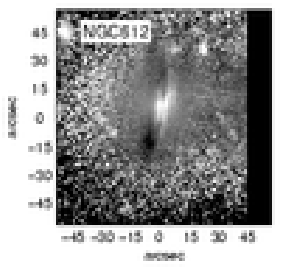} &\includegraphics[width=6cm]{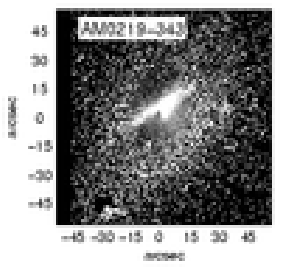}\\
  \vspace{-3mm}
 \includegraphics[width=6cm]{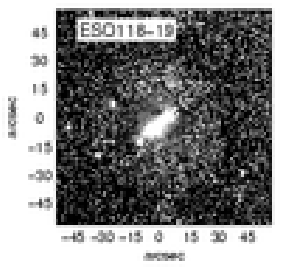} & \includegraphics[width=6cm]{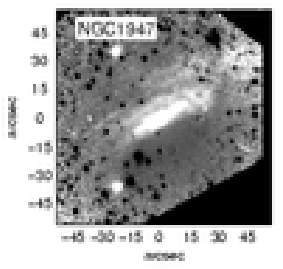} & \includegraphics[width=6cm]{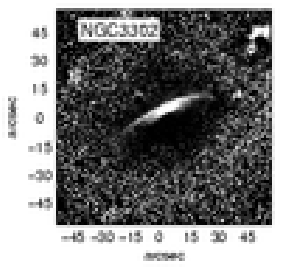}  \\
  \vspace{-3mm}
  \includegraphics[width=6cm]{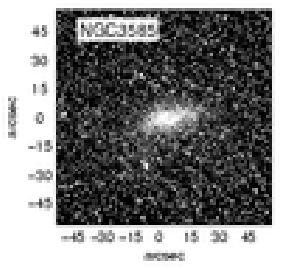} & \includegraphics[width=6cm]{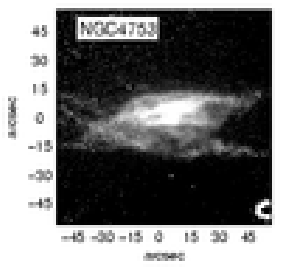} & \includegraphics[width=6cm]{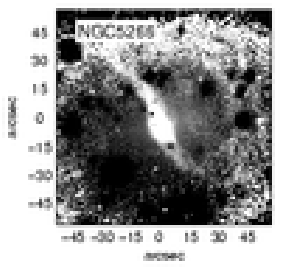}\\
 \vspace{-3mm}
\includegraphics[width=6cm]{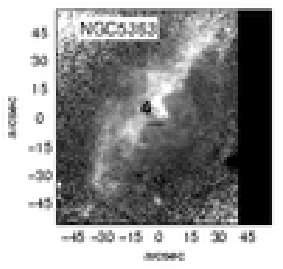} & \includegraphics[width=6cm]{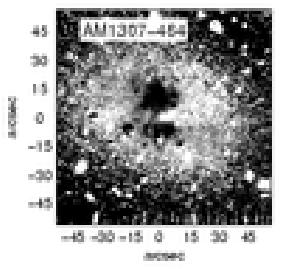} & \includegraphics[width=6cm]{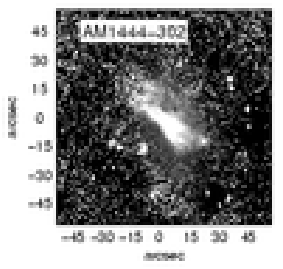} 
\end{tabular}
\end{center}
\caption{B-R colour-index maps.}
 \label{f:BRmaps}
\end{figure*}

While $\sim10-20$ exposures were obtained for each galaxy in each filter, we selected only the images with the best image quality, and with seeing below $2.5\arcsec$. The final set of images for each filter therefore included only $3-5$ images. 
Since the PSF shape of the stars in each image is often slightly distorted, we did not attempt to convolve the images to the same FWHM before combining the images. This has no consequence for the results since the spatial scale of features we measure is determined by the image with the worst PSF size.
The combined R-band contour map and B-R colour-index map for each galaxy are presented in Figures \ref{f:Rmaps} and \ref{f:BRmaps}, respectively.
%%%%%%%%%%%%%%%%%%%%%%%%%%%%%%%%%%%%%%%%%%%%%%%%%%%%%%%%%%%
\begin{table*}
 \centering
 \begin{minipage}{110mm}
  \caption{Global parameters for galaxies in our sample.
  \label{t:Obs}}
\begin{tabular}{lcclccc}
\hline
Object     & RA       & DEC       & Morph.\       & B$^0_T$ & v$_{Helio}$  & Size \\
{}         & (J2000.0)& (J2000.0) & (RC3)         &   {}    &   (km/s)     & (arcmin)\\
\hline 
AM0052-321 & 00:54:55 & -32:01:34 & -             &  14.90  &    9607      & 1.4x0.5\\
NGC612     & 01:33:19 & -36:26:38 & S0a           &  14.04  &    8995      & 1.5x1.4\\
AM0219-343 & 02:21:19 & -34:19:08 & S0            &  14.79  &    6295      & 1.4x0.9\\
ESO118-19  & 04:19:00 & -58:15:27 & S0            &  15.04  &    1239      & 0.9x0.8\\
NGC1947    & 05:26:48 & -63:45:36 & E-S0          &  11.65  &    1176      & 3.5x3.2\\
NGC3302    & 10:35:47 & -32:21:31 & S0            &  13.57  &    3794      & 1.9x1.3\\
NGC3585    & 11:13:17 & -26:45:18 & E             &  10.82  &    1434      & 6.6x3.2\\
NGC4753    & 12:52:22 & -01:11:59 & S0-a          &  10.85  &    1242      & 6.5x3.0\\
NGC5266    & 13:43:02 & -48:10:10 & S0            &  12.06  &    3707      & 3.0x2.1\\
NGC5363    & 13:56:07 & +05:15:17 & S0-a          &  11.10  &    1136      & 4.2x2.8\\
AM1307-464 & 13:10:33 & -46:59:27 & I0 pec        &  13.20  &     400      & 3.3x1.8\\
%AM1320-322 & 13:22:57 & -32:43:42 & S0            &  14.42  &    8763      & 1.3x1.2\\
%NGC5745    & 14:45:02 & -13:56:46 & S0            &  13.35  &    7085      & 1.6x0.8\\
AM1444-302 & 14:47:27 & -30:38:43 & Sa            &  14.71  &    6649      & 1.1x0.7\\
\hline
\end{tabular}
\end{minipage}
\end{table*}
%%%%%%%%%%%%%%%%%%%%%%%%%%%%%%%%%%%%%%%%%%%%%%%%%%%%%%%%%%
% %%%%%%%%%%%%%%%%%%%%%%%%%%%%%%%%%%%%%%%%%%%%%%%%%%%%%%%%%%%%%%%%%%%%%%%%%%%
% %%%%%%%%%%%%%%%%%%%%%%%%%%%%%%%%%%%%%%%%%%%%%%%%%%%%%%%%%%%%%%%%%%%%%%%%%%%
% %%%%%%%%%%%%%%%%%%%%%%%%%%%%%%%%%%%%%%%%%%%%%%%%%%%%%%%%%%%%%%%%%%%%%%%%%%%
% %%%%%%%%%%%%%%%%%%%%%%%%%%%%%%%%%%%%%%%%%%%%%%%%%%%%%%%%%%%%%%%%%%%%%%%%%%%
\begin{table*}
 \centering
 \begin{minipage}{140mm}
\caption{$R_{\lambda}$ extinction values. \label{t:Rvalues}}
\begin{tabular}{|l|cccccc|}
\hline
Object    & $R_{U1}$       & $R_{U2}$        & $R_B$            & $R_V$            & $R_R$            & $R_I$ \\
\hline
ESO118-19  & $4.84 \pm 0.60$ & $ 3.57 \pm 0.52$ & $ 3.50 \pm 0.43$ & $ 2.50 \pm 0.31$ & $ 2.02 \pm 0.25$ & $ 1.46 \pm 0.18$\\ 
NGC1947    & $4.60 \pm 0.31$ & $ 3.77 \pm 0.25$ & $ 3.53 \pm 0.23$ & $ 2.53 \pm 0.17$ & $ 1.90 \pm 0.17$ & $ 1.31 \pm 0.09$\\
NGC3302    & $4.20 \pm 0.40$ & $ 3.56 \pm 0.34$ & $ 3.52 \pm 0.33$ & $ 2.52 \pm 0.24$ & $ 1.03 \pm 0.11$ & $ 0.48 \pm 0.05$\\
NGC4753    & $5.07 \pm 0.28$ & $ 4.33 \pm 0.24$ & $ 3.96 \pm 0.22$ & $ 2.96 \pm 0.17$ & $ 2.52 \pm 0.14$ & $ 1.90 \pm 0.11$\\
NGC5266    & $5.35 \pm 0.82$ & $ 4.73 \pm 0.61$ & $ 4.41 \pm 0.57$ & $ 3.41 \pm 0.44$ & $ 2.87 \pm 0.37$ & $ 2.36 \pm 0.30$\\
NGC5363    & $4.82 \pm 0.36$ & $ 4.26 \pm 0.32$ & $ 3.98 \pm 0.29$ & $ 2.98 \pm 0.22$ & $ 2.62 \pm 0.19$ & $ 1.47 \pm 0.18$\\
AM0219-343 & $5.32 \pm 0.55$ & $ 4.17 \pm 0.43$ & $ 3.09 \pm 0.28$ & $ 2.09 \pm 0.19$ & $ 1.75 \pm 0.16$ & $ 0.76 \pm 0.10$\\
\hline 
MW galaxy  &    4.90          &   4.56           & 4.10             &3.10              & 2.32             & 1.50\\
\hline
\end{tabular}
\end{minipage}
\end{table*}
%%%%%%%%%%%%%%%%%%%%%%%%%%%%%%%%%%%%%%%%%%%%%%%%%%%%%%%%%%%%%%%%%%%%%%%%%%%
\section{Analysis}
\label{S:Analysis}
\subsection{Dust extinction law}
\label{S:dustext}
The method for deriving the extragalaxtic extinction law in the dust lanes is described shortly here and we point the reader to Finkelman et al.\ (2008) for more details.
The method assumes that elliptical galaxies have a smooth and symmetric light distribution with respect to their nuclei. It is therefore possible to fit these light distributions with a global brightness profile of a spheroid, e.g., a S\'{e}rsic profile. 
We fit ellipses to the isophotes of the galaxy. Any disturbances of the global brightness profile of the underlying galaxy can be recognized by comparing the actual observed image with the model image. Suspicious dust features are expected to cause reddening due to the extinction of light by dust. The amount of extinction can be measured by computing how much light is missing in the suspected regions relative to the fitted smooth profile.

The derivation of a dust-free model was performed in several steps. We fitted the galaxies with elliptical isophotes using the ISOPHOTE package in {\small IRAF} while masking foreground stars and determining the center of the galaxy. The fit yielded for each isophote the ellipticity, position angle and mean isophotal intensity. We created a model unextinguished galaxy from the parameters of the fitted elliptical isophotes, subtracted the observed image from the model image, and identified the dusty regions which we masked. We iterated the fitting procedure excluding all pixels with appreciable extinction ($A_V>0.02$, see e.g.\ Tran et al.\ 2001) and fixing the center if necessary. 

To measure the extinction we translated rectangular seeing-size boxes over the dust-occupied regions of each galaxy with no overlap. The amount of extinction in a particular band was then calculated on a magnitude scale by 
\begin{equation}
A_{\lambda}=-2.5 \, log\left[\frac{I_{\lambda,obs}}{I_{\lambda,mod}}\right]
\end{equation}
while excluding the nuclear regions ($\mbox{radius}\leq5\arcsec$) to avoid seeing-related effects.
No correction for Galactic extinction was made, assuming the Galactic extinction in the line of sight is uniform over each galaxy.

\begin{figure*}
\begin{minipage}{165mm}
\caption{Extinction curves for the program galaxies (solid lines) along with the canonical curve for the Galaxy (dashed lines) for comparison. Extinction values derived in Finkelman et al.\ (2010) are represented by open circles. The error bars are $1\sigma$ errors. \label{fig:curves}}
\includegraphics[trim=20mm 0mm 0mm 0mm, clip]{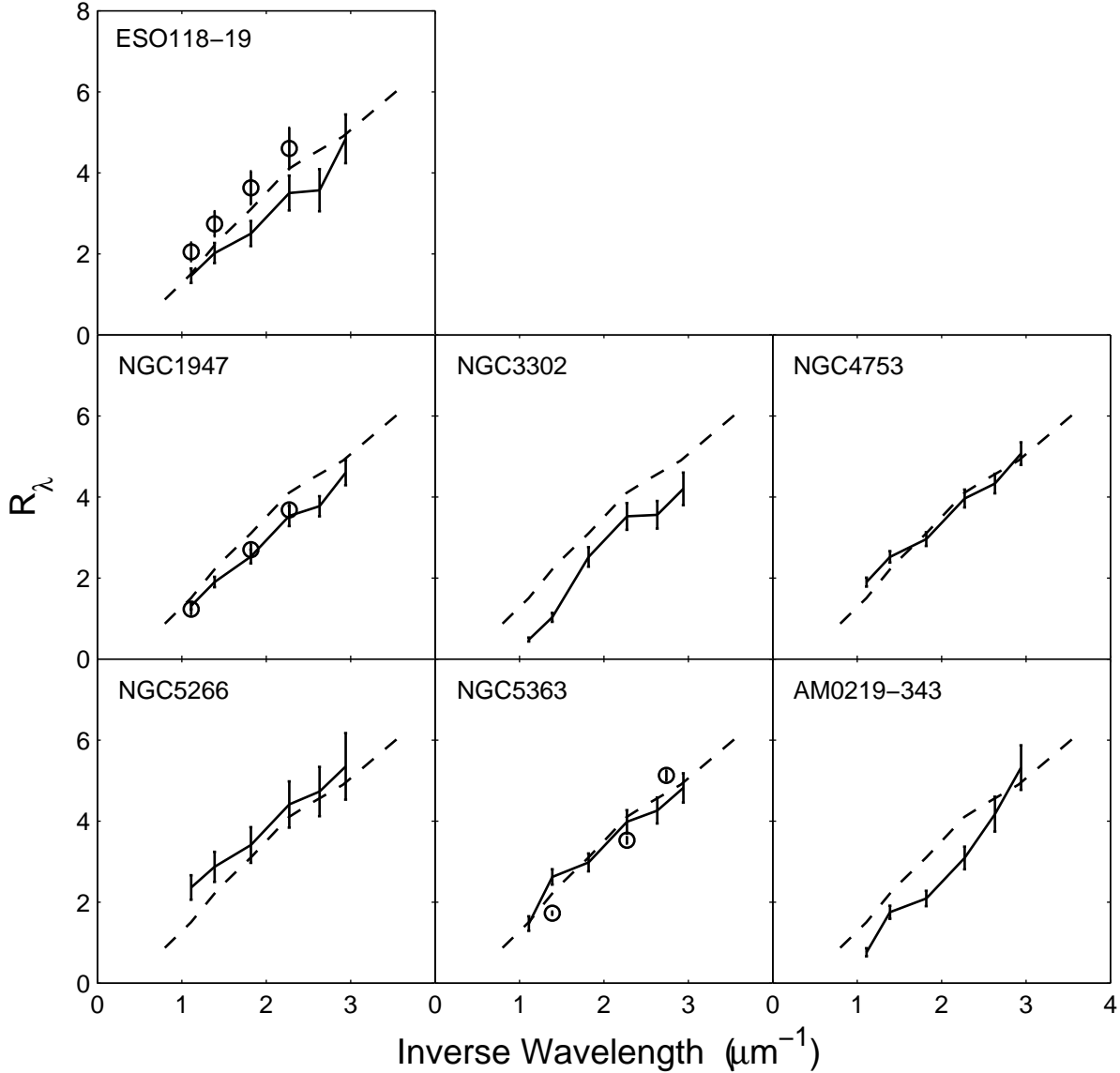}
\end{minipage}\end{figure*}

By fitting a linear regression between the total extinction $A_\lambda$ at wavelength $\lambda$ and the V-band extinction $A_V$  measured along the dusty region in different bands we derived the extinction values for each galaxy. The best-fitting slopes were then used to calculate the $R_{\lambda}\left( {\equiv}\frac{A_\lambda}{E\left( B-V\right)} \right)$ values for each galaxy and were compared with the Galactic values. 
The obtained $R_\lambda$ extinction values are listed in Table \ref{t:Rvalues}, along with the canonical extinction values taken from Savage \& Mathis (1979) for comparison. The extinction curves are plotted in Figure \ref{fig:curves}. 

This method is limited to galaxies where the dust covering factor is relatively small, and requires that the dusty regions are well-sampled. Therefore, for five galaxies where the apparent dust coverage is restricted to the inner parts of the galaxies or where the S\'{e}rsic profile fits badly the galaxy we could not derive an extinction law as further described below.

AM0052-321 shows a prominent dust lane along its major axis, but reveals a shell structure which prevents a proper modeling of the underlying galaxy. 
%We also note that while AM0052-321 was described by Arp \& Madore (1987) as one object in an interacting pair of two elliptical-like galaxies (Arp et al.\ 1987), the second galaxy clearly reveals stellar spiral arms in our images. 
The optical image of the E/S0 galaxy AM1307-464 shows a patchy dust structure in its inner regions. However, the B-R colour image shows that this region is bluer than the underlying galaxy, implying that the colour variation is induced by a recent star formation episode and not by the presence of dust (see also Davidge 2007).
Although AM1444-302 is classified Sa, it appears in our images as an elliptical galaxy with a complex structure of dust lanes in its inner part. Since most of the inner region is obscured, and since the outer regions show shells, the underlying galaxy cannot be modelled.
The B-R colour image of NGC3585 shows that the inner disk is redder with respect to the bulge (see also Fisher et al.\ 1996), implying the presence of dust. Since the inner isophotes of NGC3585 are disky, suggesting the presence of a inner disk within the dominant bulge, we cannot model the underlying galaxy in that region by using our method. In addition, our images show that the dust lane across the bulge of NGC612 actually follows a stellar disk at high inclination, resembling the well-known Sombrero galaxy. %%%%%%%%%%%%%%%%%%%%%%%%%%%%%%%%%%%%%%%%%%%%%%%%%%%%%%%%%%%%%%%%%%%%%%%%%%%
%%%%%%%%%%%%%%%%%%%%%%%%%%%%%%%%%%%%%%%%%%%%%%%%%%%%%%%%%%%%%%%%%%%%%%%%%%%
\subsection{Dust mass estimation}
\label{S:dustmass}
Assuming the extragalactic dust grains are uniform throughout each galaxy and are similar to those in the MW we use the Mathis, Rumpl \& Nordsieck (1977) two-component model consisting of individual spherical silicate and graphite grains with an adequate mixture of sizes to calculate dust masses.
The dust mass is then calculated by integrating the dust column density over the image areas S occupied by dust lanes, yielding:
\begin{equation}
M_d=\mbox{S}{\times}\Sigma_d=\mbox{S}{\times}l_d{\times}\int\limits_{a_-}^{a_+}\frac{4}{3}{\pi}a^3{\rho}_dn\left(a\right)da .
\end{equation}
where $a$ is the grain radius; $n\left( a\right)$ represents the size distribution of dust grains; $a_-$ and $a_+$ represent the lower and upper cutoffs of the size distribution, respectively; ${\rho}_d$ is the specific grain densities which is taken to be $\sim 3$ gr cm$^{-3}$ for graphite and silicate grains (Draine \& Lee 1984); $l_d$ represents the dust column length along the line of sight; and $n_H$ is the hydrogen number density (see Finkelman et al.\ 2008).

The value of the product $l_d \times n_H$ is inferred by measuring total extinction and calculating the extinction efficiency (Finkelman et al.\ 2008). The hidden assumption is that the dust forms a foreground screen for the galaxy and neglecting any intermix of dust and stars within the galaxy. The dust mass estimate from total extinction values provides therefore only a lower limit to the real dust content of the host galaxies. 

Alternatively, the dust mass can be estimated based on the observed far-IR dust emission assuming thermal equilibrium. The relation between the dust mass and the observed far-IR flux density, $I_{\nu}$, is 
\begin{equation}
M_d=\frac{4}{3} a \rho _d D^2 \frac{I_{\nu}}{Q_{\nu}B_{\nu}(T_d)}
\end{equation}
where ${\rho}_d$ and $D$ are the specific grain mass density and distance of the galaxy in Mpc, respectively; $Q_{\nu}$ and $B_{\nu}(T_d)$ are the grain emissivity and the Planck function for the temperature $T_d$ at frequency $\nu$, respectively (see Hildebrand 1983).
The dust grain temperature can be calculated from the IRAS flux densities at 60 $\mu$m and 100 $\mu$m using $T_d=\left( \frac{S_{60}}{S_{100}}\right)^{0.4}$  (Young et al.\ 1989). Hot circumstellar dust may also contribute to the 60 $\mu$m and 100 $\mu$m passbands, although this effect on the 60 $\mu$m and 100 $\mu$m flux densities is only a few percent (Goudfrooij \& de Jong 1995).
The quantity $a{\rho}_d/Q_{\nu}$ can be calculated for silicate and graphite grains assuming it is independent of $a$ for  $\lambda \gg a$ (see Hildebrand 1983; Finkelman et al.\ 2008).

The calculated dust mass estimates represent lower limits, since IRAS was not sensitive to dust cooler than about 20K, which emits mostly at $\lambda>100$ $\mu$m.
Table \ref{t:dmass} lists the estimated dust mass from the total optical extinction and the estimated dust mass based on IRAS flux densities taken from the catalog of Knapp et al.\ (1989) for bright early-type galaxies. We assume for the calculations a standard cosmology with H$_0$=73 km sec$^{-1}$ Mpc$^{-1}$.

\begin{table*}
 \centering
\caption{Dust mass. \label{t:dmass}}
\begin{tabular}{lcccc}
\hline
Object     & \multicolumn{2}{c}{IRAS flux (mJy)} &  $ \mbox{Log} \left(\frac{M_d}{M_ \odot}\right)_{\mbox{IRAS}}$ & $ \mbox{Log}\left( \frac{M_d}{M_ \odot}\right) _{\mbox{optical}} $  \\
   {}       & 60$\mu$m      & 100$\mu$m      &     {}           &    {}           \\
\hline
ESO118-19   & $733  \pm 51$ & $1391 \pm 84 $ & $4.70\pm0.08$    & $4.44 \pm 0.02$ \\
NGC1947	    & $1052  \pm 53$& $4136 \pm 248$ & $5.69\pm0.09$    & $5.01 \pm 0.01$ \\
NGC3302     & $1876 \pm206$ & $5401\pm648$   & $6.56\pm0.17$    & $5.55 \pm 0.02$ \\ 
NGC4753     & $2438 \pm 293$& $9008 \pm 1171$& $6.02\pm0.20$    & $5.20 \pm 0.01$ \\
NGC5266     & $1220 \pm 98$ & $4365\pm349$   & $6.63\pm0.12$    & $5.04 \pm 0.03$ \\ 
NGC5363     & $1693 \pm 169$& $5150 \pm 567$ & $5.54\pm0.16$    & $4.86 \pm 0.02$ \\
AM0219-343* & $0 \pm 42$    & $0 \pm 210$    & $<8.2$           & $6.00 \pm 0.02$ \\
\hline
\end{tabular}
\label{t:DMass}
\begin{minipage}[]{13cm}
\begin{small}
*Upper limits assuming $\mbox{T}=20\mbox{K}$
\end{small}
\end{minipage}
\end{table*}
%%%%%%%%%%%%%%%%%%%%%%%%%%%%%%%%%%%%%%%%%%%%%%%%
\subsection{Fine dust structure - the case of NGC2685}
The derivation of an extinction law is restricted by the image angular scale and therefore requires measuring a mean extinction value for each resolution element. However, the presence of dust-free ``holes'' or high-opacity ``bricks'' in the dust structure can significantly affect these measurements (see for example Keel \& White 2001).

To quantify the effects of dust on scales smaller than the $\sim0.1-1$ kpc resolution provided by SALTICAM we use data from HST/WFPC2. Unfortunately HST did not observe any of our sample galaxies in the optical region; we therefore examine NGC2685, a well-known dust-lane galaxy for which HST data is available, as a representative example.
NGC2685 was observed with WFPC2 in the B (F450W), V (F555W) and I (F814W) passbands. The total exposures were 2300 sec (B), 1000 sec (V) and 730 sec (I), split into halves. The individual exposures in each filter were combined using the CRREJ task in STSDAS for cosmic-ray recognition and rejection. 

\begin{figure}
%\begin{minipage}{165mm}
\caption{Distribution of B-band transmission by area. The upper panel shows the differential distribution of dusty regions with different optical depth values while translating a $2\times2$ pixel box ($\sim$5 pc; solid line) and a $16\times16$ pixel box ($\sim$40 pc; dashed line) along the dust lanes. The data suggest that most regions have B transmission from 0.2 to 0.9. The lower panel refers to the cumulative distribution of the data presented in the upper panel.  \label{f:histogram}}
\includegraphics[trim=0mm 30mm 50mm 0mm, clip]{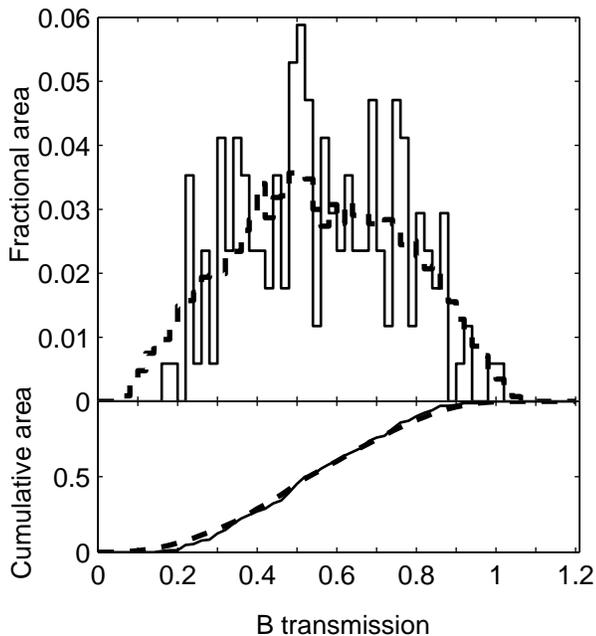}
%\end{minipage}
\end{figure}

While on $\sim$kpc scales the dust in NGC2685 seems to be organized in a well-defined helical structure, HST/WFPC2 images clearly reveal more complicated structures on much smaller scales. 
The $\sim0.1\arcsec$ PSF resolution provided by HST/WFPC2 allows deriving the extinction down to $2\times2$ pixel boxes, and therefore studying the extinction on a spatial scale of $\sim5$ pc. We plot in Figure \ref{f:histogram} the relative distribution of the B transmission values (${\equiv}10^{-A_{B}}$) as measured for $2\times2$ and $16\times16$ pixel boxes translated along the dust lane. Comparing the two relative distributions and the computed cumulative distributions hints at a rather smooth dust coverage and supports a scale-free structure on these scales. 
Furthermore, the B transmission values in the sub-structures vary from 0.2 to 0.9 and are consistent with low opacity within most of the obscured area (see for comparison Figure 6 in Keel \& White 2001).

The scale-free dust distribution can also be examined in terms of fractal analysis (see Mandelbrot 1982). 
A similar analysis of the interstellar dust structure in several nearby spiral galaxies revealed a hierarchical, fractal-like structure in a wide range of scales from $\sim0.1$ pc to $\sim1$ kpc (e.g., Westpfahl et al.\ 1999; Keel \& White 2001; Sanchez \& Alfaro 2008).
Simply put, a fractal structure possesses details at every scale and a fractal dimension determines whether a system is homogeneous, and what fraction of space is filled (Combes 2000; S\'{a}nchez, Alfaro \& P\'{e}rez 2005).
Astrophysical fractals are an approximation in a range between two limiting scales and are different from pure mathematical fractals that are infinite (Combes 2000). 

One method to test structural scaling or fractal behavior is by constructing a scale-free relation between the perimeter of a contour and the scale length over which it is integrated or smoothed (Westpfahl et al.\ 1999; Keel \& White 2001; Chappell \& Scalo 2001). The box-counting procedure provides a simple perimeter-scale test (Westpfahl et al.\ 1999). The scale-free relation, also known as the box-counting dimension, can be obtained by a power-law fit to the distribution of the number of boxes for different box sizes. Linear structures show a slope of $-1$, while fractal structures show a slope between $-1$ and $-2$. The box-counting dimension is the negative of this slope.

\begin{figure}
%\begin{minipage}{165mm}
\caption{Box counting analysis of the relation between length (region perimeter) and smoothing length for a $256 \times 128$ pixel$^2$ region in NGC2685. Single contours were evaluated for total extinction values from 0.5 to 1.5 represented in the plot by a different symbol with the lines marked by the extinction value. Scale-free (fractal) behavior is manifested as straight lines in this log-log plot.  \label{f:fractal}}
\includegraphics[]{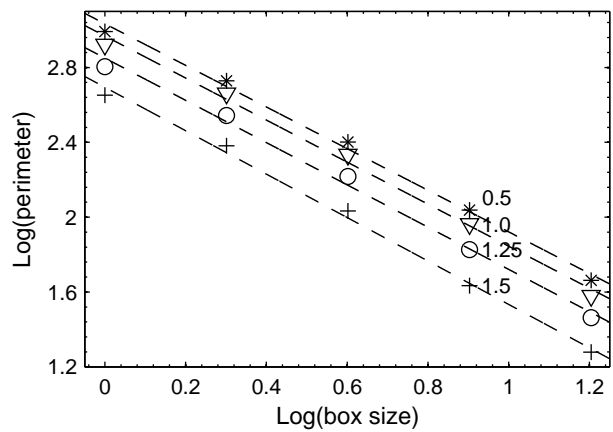}
%\end{minipage}
\end{figure}

Following the box-counting algorithm we used the extinction maps to evaluate the length of single-total extinction contours in the range $A_B= 0.5-1.5$ for smoothing lengths of 2, 4, 8 and 16 pixels, so that no resampling of pixel values affects the results. 
Figure \ref{f:fractal} plots on a log-log scale the perimeter calculated for each contour for each box size. The self-similar behavior of each contour is manifested as a straight line in this log-log plot, with slopes in the range $\sim(-1.1)-(-1.2)$ derived for the different contours. 

\begin{figure}
\caption{Extinction curve for NGC2685 from HST (upper curve) and SDSS (lower curve) data. The extinction values (plus signs) are presented along with the canonical curve for the Galaxy (dashed lines) for comparison. The error bars are $1\sigma$ errors.  \label{f:helixextlaw}}
\begin{center}
\includegraphics[trim=0mm 0mm 5cm 0mm, clip]{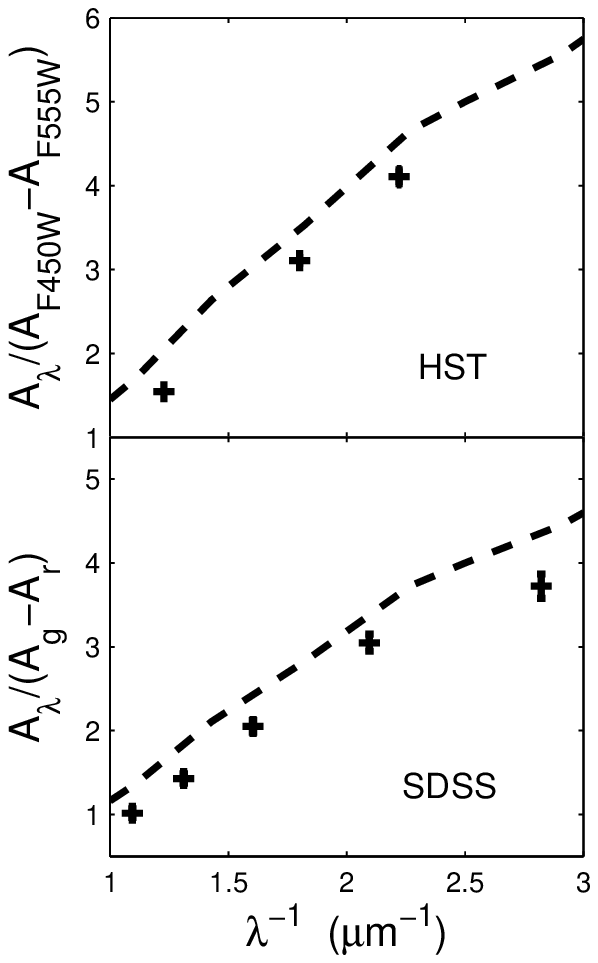} 
\end{center}
\end{figure}

Determining the extinction law of a self-similar dust distribution is not expected to be affected by the linear scale on which the extinction is measured. 
To verify this we derived the extinction law by measuring the extinguished light within $2\times2$ and $16\times16$ pixel boxes. The results are found to be consistent within the error bars and the mean extinction values are plotted in Figure \ref{f:helixextlaw}. For comparison, we plot in Figure \ref{f:helixextlaw} also the extinction values measured by using the same method on SDSS images with $\sim1.2$ arcsec resolution. 
%%%%%%%%%%%%%%%%%%%%%%%%%%%%%%%%%%%%%%%%%%%%%%%%%%%%%%%%%%%%%%%%%%%%%%%%%%%
%%%%%%%%%%%%%%%%%%%%%%%%%%%%%%%%%%%%%%%%%%%%%%%%%%%%%%%%%%%%%%%%%%%%%%%%%%%
%%%%%%%%%%%%%%%%%%%%%%%%%%%%%%%%%%%%%%%%%%%%%%%%%%%%%%%%%%%%%%%%%%%%%%%%%%%
\section{Discussion}
\label{S:Discussion}
\subsection{Dust extinction}
We derived extinction curves and extinction values for seven of our 12 sample galaxies, excluding AM0052-321, AM0131-365, NGC3585, AM1307-464 and AM1444-302.
The mean extinction value for the galaxies, $R_V=2.71\pm0.43$, matches the $\sim2.80$ value derived in previous studies (e.g., Patil et al.\ 2007; Finkelman et al.\ 2008) for E/S0 galaxies with well-settled dust lanes and is slightly smaller than that in the MW, but similar within the error bars. These $R_V$ values are also within the range of values derived by Goudfrooij et al.\ (1994), 2.1 to 3.3, which implies a characteristic dust grain size up to 50\% smaller than that of standard Galactic dust grains (see Finkelman et al.\ 2008). 
The lowest extinction value measured in a galaxy in our sample, $R_V=2.09 \pm 0.19$, is similar to the lowest extinction values obtained by Goudfrooij et al.\ (1994), Patil et al.\ (2007) and Finkelman et al.\ (2008). 
These results further support Goudfrooij et al.\ (1994) conclusion that E/S0 galaxies with smooth and well-defined dust lanes exhibit smaller $R_V$ values (see also Patil et al.\ 2007; Finkelman et al.\ 2010).

We show that although the derived extragalactic extinction curves are close to the standard Galactic value, most of them lie above or below the Galactic extinction curve and parallel to it. 
Deviations of extinction values from the Galactic values are generally linked with a variation of the characteristic grain size responsible for the optical extinction in such galaxies with respect to dust in the MW (see Finkelman et al.\ 2008 and references therein). However, it should first be established that the measured extinction law represents the ``true'' extinction law, as discussed below. 

\subsection{Effect of young stars}
Goudfrooij et al.\ (1994) discussed the effect star formation and light emission from young massive stars on the derived extinction law.
Finkelman et al.\ (2010) studied a sample of galaxies where the dusty structure is generally morphologically associated with ionized gas; this hints at the presence of a small fraction of young stars and of a recent star formation episode.
Since the light of the underlying galaxy is extrapolated from the unobscured regions, and since the presence of a young stellar population in the dusty locations is expected to be bluer than the old stellar population, the extinction might be underestimated, especially at the shorter wavelengths. A uniform young stellar population in the dust lane will have the same effect over all obscured regions and therefore should not affect the slope of the $A_\lambda$ vs. $A_V$ relation, but only its intercepts with the axes. However, with a more complex stellar structure the real extinction curve may be more concave than the measured one and show larger extinction values.

The effect of forward scattering in dusty regions on the observed extinction curve can be neglected as well. This effect is mainly dominant at the edges of disks and leads to apparently lower extinction values in the blue and to a relatively flat extinction curve with respect to the Galactic curve, whereas such flattening is not evident in any of our sample galaxies (see Finkelman et al.\ 2008). 

\subsection{Effect of resolution and scales of dust distribution}
The PSF of the SALT observations is typically 2.5\arcsec. The extinction is therefore measured with an assumption that 
the dust structure is homogeneous and smooth within each resolution element. Furthermore, the hidden assumption is that $A_\lambda=1.086 \, \tau_\lambda$ is valid for a given wavelength only as long as the dust is a foreground screen, while the optical depth $\tau_\lambda$ may vary with column density and column length. 
Finkelman et al.\ (2008) studied the influence of the dust location on the observed extinction law for the case of varying optical depths concluding that if the dusty locations are optically-thin, i.e., $\tau_V\simeq1$ as typical for our sample galaxies, the deviation from the ``real'' linear relation may not be significant. 
%%%%%%%%%%%%%%%%%%%%%%%%%%%%%%%%%%%%%%%%%%%%%%%%%%%%%%%%%%%%%%%%%%%%%%%%%%%
\begin{figure*}
\caption{Left panel: Extinction values measured in NGC1947 from SALT data. Right panel: Extinction values calculated from HST/WFPC2 by measuring the intensity in 2x2 (grey plus signs) and 16x16 pixel boxes (black plus signs) translated across the dusty regions of NGC2685.   The solid lines represent the model best-fit assuming dust is a foreground screen. \label{f:extval}}
\begin{center}
\begin{tabular}{cc}
 \includegraphics[trim=0mm 10mm 0mm 10mm, clip, width=8cm]{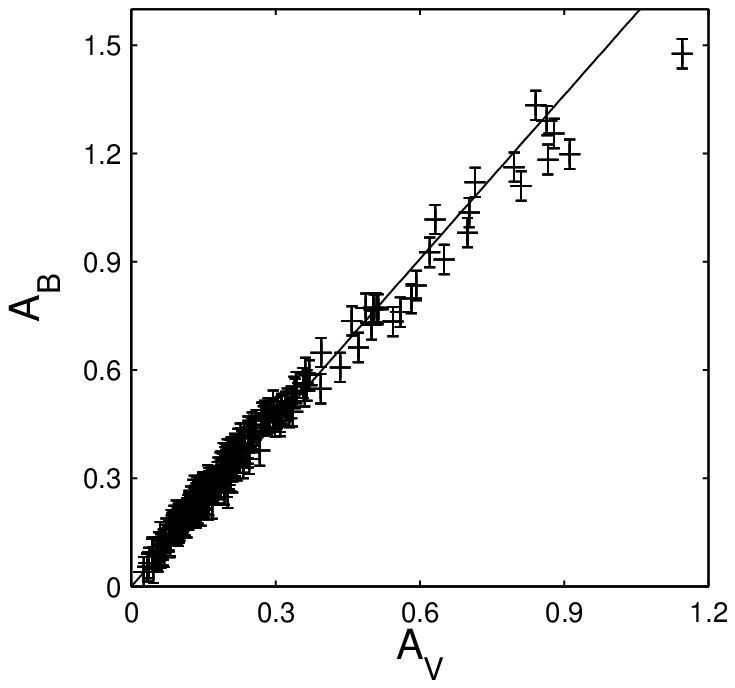} & \includegraphics[trim=0mm 10mm 0mm 10mm, clip, width=8cm]{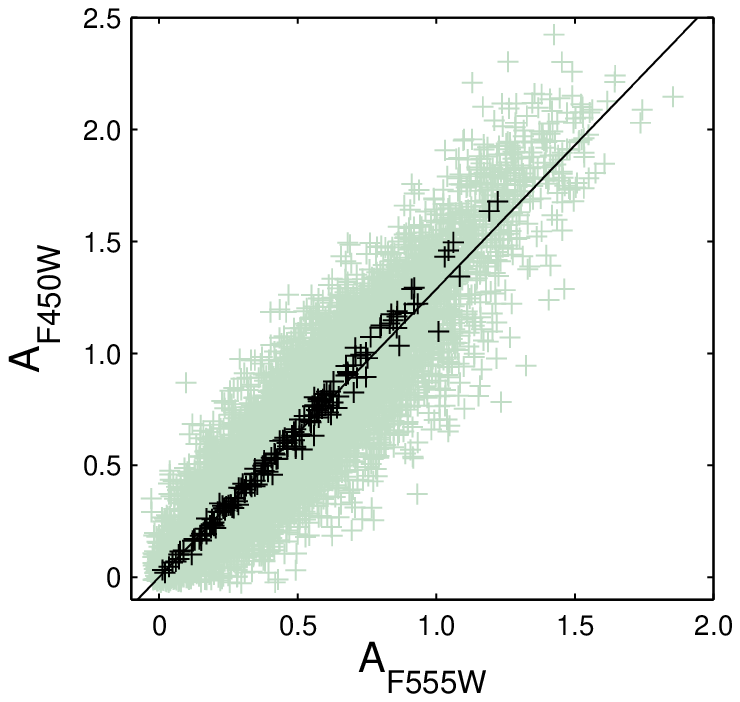}
\end{tabular}
\end{center}
\end{figure*}
%%%%%%%%%%%%%%%%%%%%%%%%%%%%%%%%%%%%%%%%%%%%%%%%%%%%%%%%%%%%%%%%%%%%%%%%%%%
However, tracing the possible variations of the dust column density and geometry within the dust lane in a certain galaxy requires better spatial resolution than provided by SALTICAM. With no available high-resolution optical data of our sample galaxies, we used archival HST/WFPC2 data of NGC2685, a representative E/S0 galaxy with well-defined dust lanes, to test the Finkelman et al.\ (2008) hypothesis. In particular, we investigated the presence of highly obscured regions through the search of greyish extinction. Our analysis shows that the dust lanes do not show transparent ``holes'' or opaque ``bricks'', but are rather smooth structures down to scales as small as $5-10$ pc. 
Figure \ref{f:extval} further demonstrates the foreground screen model validity for the case of our galaxies. 

The dust structure in NGC2685 is characterized by a fragmented and filamentary geometry. The complexity of structures of interstellar clouds are often studied and characterized by fractal analysis (see Khalil et al.\ 2006; S\'{a}nchez \& Alfaro 2008 and references therein). The limited resolution, finite-size effects  and use of different analysis algorithms are known to introduce biases on the estimate of the fractal dimension. Using the box-counting technique, previous studies of molecular and dust clouds found that the fractal dimension value is almost always between 1.2 and 1.5. 

Although a significant part of fractal analysis studies is dedicated to the distribution of gas and dust in the Galaxy (e.g., Khalil et al.\ 2006), a growing number of studies also focus on the HI (e.g., Westpfaul et al.\ 1999), HII (e.g., S\'{a}nchez \& Alfaro 2008) and the dust component (e.g., Keel \& White 2001; Elmegreen, Elmegreen \& Leitner 2003) in nearby spiral galaxies.  Therefore, the general result of a more or less constant fractal dimension might reflect a universal value that is related with the underlying physical processes that drive the growth and subsequent evolution of the interstellar clouds (e.g., Elmegreen \& Falgarone 1996; Elmegreen 1997; Combes 2000; Chappell \& Scalo 2001). The similarity in fractal structure of both dust and molecular clouds also support an hypothesis that the dust is related with the process of molecule formation in cold clouds (Datta 2004).

Examining the extinction map of NGC2685 we find the contours at constant extinction to be individually scale-free for sampling lengths $\ge2$ pixels ($\ge5$ pc). We measured for NGC2685 a box-counting dimension between 1.1 and 1.2, which is only marginally consistent with the fractal dimension measured for the ISM in previous studies, and is closer to the fractal dimension of linear and euclidean structures. 
Therefore, it is not clear whether the fractal behavior of ISM in dust-lane E/S0 galaxies follows the universal structural evolution of ISM observed in the MW and nearby spiral galaxies or otherwise indicates a different evolution. Determining this with confidence requires measuring with HST the fractal dimension of the dust structures in a larger sample of dust-lane E/S0s.

\subsection{Dust mass}

The dust mass for each sample galaxy was evaluated independently in two ways, based on the optical extinction or from far-IR observations. The estimated masses lie in the range $\sim 10^4$ to $10^7M_{\odot}$, in good agreement with previous estimates for early-type galaxies with dark lanes (see Finkelman et al.\ 2008 and references therein). 
Since we could not measure the entire dusty region with the optical method, and since we assumed that the dust does not intermix with the stars within the galaxy, this method provides only a lower limit to the true dust content of the host galaxies that should be confirmed by other methods. Using IRAS flux densities to trace the dust far-IR emission reveals that the dust masses are up to an order of magnitude higher than measured from the optical extinction, implying the presence of a more massive diffuse dust component (Goudfrooij \& de Jong 1995; Patil et al.\ 2007; Finkelman et al.\ 2008). 

\section{Conclusions}
\label{S:summ}
This is the second paper presenting new results from an on-going program to study the extragalactic extinction law with SALT down to the ultraviolet atmospheric cutoff (see Finkelman et al.\ 2008). 

We derived the ratio between the total V band extinction and the selective B and V extinction $R_V$ for seven out of our 12 sample galaxies, and obtained an avergage of $2.71\pm0.43$, slightly smaller than that found by Finkelman et al.\ (2008) and than the Galactic value but similar within the error bars.
Extinction curves derived for the sample galaxies run parallel to the canonical Galactic curve but do not necessarily overlap.
We found that deviations from the standard Galactic value are unlikely to be affected by the image resolution and scales of the dust distribution, and that the effect of intermix of stars and dust, which needs to be considered when measuring the extinction in highly-inclined disk galaxies, is negligible for the case of dust lanes in E/S0 galaxies. 
We conclude that dust in the extragalactic environment has similar properties to the Galactic grains, and that any deviation of extinction value from the Galactic one can be accounted for by a variation of the characteristic grain size responsible for the optical extinction.

The dust content in the dust lanes derived by the optical extinction method for the sample galaxies is in the range $\sim 10^4$ to $10^7M_{\odot}$, whereas the diffuse dust content might be larger by an order of magnitude. These results are fully in agreement with those of Finkelman et al.\ (2008).

\subsection*{Acknowledgments}
We thank the referee, Madhav Patil, for providing constructive comments and help improving the content of this paper.

Some of the observations reported in this paper were obtained with the Southern African Large Telescope (SALT).

Based on observations made with the NASA/ESA Hubble Space Telescope, obtained from the data archive at the Space Telescope Institute. STScI is operated by the association of Universities for Research in Astronomy, Inc. under the NASA contract  NAS 5-26555. 

Funding for the SDSS and SDSS-II has been provided by the Alfred P. Sloan Foundation, the Participating Institutions, the National Science Foundation, the U.S. Department of Energy, the National Aeronautics and Space Administration, the Japanese Monbukagakusho, the Max Planck Society, and the Higher Education Funding Council for England. The SDSS Web Site is http://www.sdss.org/.

The SDSS is managed by the Astrophysical Research Consortium for the Participating Institutions. The Participating Institutions are the American Museum of Natural History, Astrophysical Institute Potsdam, University of Basel, University of Cambridge, Case Western Reserve University, University of Chicago, Drexel University, Fermilab, the Institute for Advanced Study, the Japan Participation Group, Johns Hopkins University, the Joint Institute for Nuclear Astrophysics, the Kavli Institute for Particle Astrophysics and Cosmology, the Korean Scientist Group, the Chinese Academy of Sciences (LAMOST), Los Alamos National Laboratory, the Max-Planck-Institute for Astronomy (MPIA), the Max-Planck-Institute for Astrophysics (MPA), New Mexico State University, Ohio State University, University of Pittsburgh, University of Portsmouth, Princeton University, the United States Naval Observatory, and the University of Washington.

\end{document}